%% file: myglauber.tex
\documentclass[conference]{IEEEtran}
\makeatletter
\def\ps@headings{%
\def\@oddhead{\mbox{}\scriptsize\rightmark \hfil \thepage}%
\def\@evenhead{\scriptsize\thepage \hfil \leftmark\mbox{}}%
\def\@oddfoot{}%
\def\@evenfoot{}}
\makeatother
\usepackage{url}

\usepackage[usenames,dvipsnames]{pstricks}

\usepackage{mdwmath}
\usepackage{mdwtab}
\usepackage{graphicx}

\usepackage[english]{babel}
\usepackage[T1]{fontenc}
\usepackage{amsmath,amssymb,bbm,amsthm}
\usepackage{enumerate}
\usepackage{pst-3dplot,pst-grad}

\begin{document}
\input{macro}
%
\title{Modeling energy consumption in cellular networks}

\author{\IEEEauthorblockN{L. Decreusefond}
  \IEEEauthorblockA{Telecom Paristech, LTCI\\
    Paris, France
  } \and \IEEEauthorblockN{T.T. Vu}
  \IEEEauthorblockA{Telecom Paristech, LTCI\\
    Paris, France} \and \IEEEauthorblockN{P. Martins}
  \IEEEauthorblockA{Telecom Paristech, LTCI\\
    Paris, France}}

\maketitle

\begin{abstract}
  In this paper we present a new analysis of energy consumption in
  cellular networks. We focus on the distribution of energy consumed
  by a base station for one isolated cell. We first define the energy
  consumption model in which the consumed energy is divided into two
  parts: The additive part and the broadcast part. The broadcast part
  is the part of energy which is oblivious of the number of mobile
  stations but depends on the farthest terminal, for instance, the
  energy effort necessary to maintain the beacon signal. The additive
  part is due to the communication power which depends on both the
  positions, mobility and activity of all the users.  We evaluate by
  closed form expressions the mean and variance of the consumed
  energy. Our analytic evaluation is based on the hypothesis that
  mobiles are distributed according to a Poisson point process. We
  show that the two parts of energy are of the same order of magnitude
  and that substantial gain can be obtained by power control. We then
  consider the impact of mobility on the energy consumption. We apply
  our model to two case studies: The first one is to optimize the cell
  radius from the energetic point of view, the second one is to
  dimension the battery of a base station in sites that do not have
  access to permanent power supply.
\end{abstract}


%
\IEEEpeerreviewmaketitle

\section{Introduction}


According to the GSM Association, \textsl{more than 80$\%$ of a
  typical mobile network operator's energy requirements are associated
  with operating the network. The typical annual CO2 emissions per
  average GSM subscriber is now about 25kg CO2, which equates to the
  same emissions created by driving an average European car on the
  motorway for around one hour.  However, the mobile industry
  continues to look for ways to reduce energy needs. Air conditioning
  is being replaced by fans or passive air flows whenever
  possible. Several programs are aiming to deploy solar, wind, or
  sustainable bio fuels technologies to 118,000 new and existing
  off-grid base stations in developing countries by 2012.  Network
  optimization upgrades currently can reduce energy consumption by
  44$\%$ and solar-powered base stations could reduce carbon emissions
  by 80$\%$. Optimization of the physical network through improved
  planning and the spectrum allocations for mobile broadband can also
  contribute to significant energy savings}.

As a consequence of the previous statements, it appears clearly that
energy consumption must be taken into consideration at the very
beginning of the conception of cellular networks. For the development
of cellular communications in emerging countries, it is necessary to
be as energy conservative as possible by using the least possible
number of base stations for a given quality of service. As the size of
a cell covered by a given base station depends essentially on the
emitting power of its antenna, the smaller the size of a cell, the
less the consumed energy. However, when base stations cover a small
region, many of them are necessary to cover a given region. There is
thus a trade-off between the number and the coverage of each base
stations. In order to fix the optimal radius of a cell, one must have
quantitative models of the energy consumed at a base station in terms
of positions, locations and traffic of the terminals. Furthermore, if
we think about the deployment of base stations in low populated
regions without power supply, base stations should be energetically
autonomous, thus powered by a battery, be it solar or chemical. The
energy consumed in such a situation is thus a key parameter in the
building of a cellular network. These are the two questions we aim to
answer in the following considerations.

Several measurement based analysis pointed out the different aspects
in energy consumption (see
\cite{fehskefootprint,FeZim08EnergyCons,WangVasila2011SurvGreen,4604612}
and references therein). In \cite{ullah_applicability_2012}, some
models are proposed for the activity of a single terminal and the
resulting energy consumption. In
\cite{chinara_mobility_2008,heinzelman_energy-efficient_2000}, the
choice of the cluster heads in Ad-Hoc networks integrates energy
consumption consideration and take into account the geometry of the
terminals by considering the proximity graph. In
\cite{wu_minimum-energy_2005}, energy saving motivates to use network
coding in ad-hoc networks. As a conclusion, there are many
investigations about how to save energy but no model seems to emerge
in order to evaluate quantitatively the energy consumption. There is
however one notable exception which is the paper
\cite{xiang_energy_2013}. In that work, a stochastic geometry based
model for energy consumption in a cellular network is considered. It
assumes that mobiles are connected to their closest base stations and
introduces an energy consumption model based on the distance between
base stations and mobiles, taking into account interference due to the
presence of several base stations. We go further in this direction by
considering a refined model for the energy consumption in an isolated
cell, as it includes the energy devoted to broadcast messages (like
the beacon signal) and takes into account both traffic activity and
users mobility. Moreover, instead of relying on simulation results, we
give as much as possible closed form formulas for different statistics
of the energy consumption. This leads to qualitative results showing
the importance of the path-loss exponent (see below for its
definition).

This paper is organized as follows: In
Section~\ref{sec:primer-poisson-point}, we recollect basic and
advanced facts about Poisson point processes in general spaces. We
then present the system model based on a Poisson point process
including not only the positions but also the traffic activity and the
mobility pattern of each user.  In Section~\ref{sec:motionless-users},
we first evaluate the consumed energy for motionless users. We show,
in Section~\ref{section: Onoffmobility}, that mobility does not change
the mean value but decreases the variance of the consumed
energy. Thus, as far as dimensioning is concerned, it is conservative
to consider that users do not move. We apply these considerations in
Section~\ref{sec: Application} to two case studies: Finding the
optimal radius of a cell under energy constraints and estimating the
power of a battery to maintain a functioning network during a
given~time.

\section{A primer on Poisson point process}
\label{sec:primer-poisson-point}

A Poisson process $N$ on the real line admits a usual description
based on a sequence $(S_n,\, n\ge 1)$ of independent exponentially
distributed random variables. Denote by $\lambda$ the parameter of
$S_1$. The atoms of a Poisson process are the sequence
$(T_n=S_1+\ldots+S_n,\, n\ge 1)$. Then, one can prove
\cite{Decreusefond:2012sys} that the number of points in a domain of
Lebesgue measure $l$ is a Poisson random variable of parameter
$\lambda l$. Moreover, given $N(D)=n$, i.e. given the number of points
in $D$ is equal to $n$, the atoms of $N$ are independently and
uniformly distributed over $D$. This explains the definition of a
Poisson process in any dimension.
\begin{definition}
  $\pi_\mu$ a measure on the set configuration on $E=\R^d$, is a
  Poisson point process (PPP) of intensity $\nu$ if for all sets
  $(C_1,\cdots,\,C_n)$ of mutually disjoint compact subsets of $E$:
  \begin{multline}\label{eq_myglauber:4}
    \Pc(N(C_1)=k_1,\cdots,\,N(C_n)=k_n) \\ =
    \prod_{i=1}^n\left(e^{-\nu(C_i)}\frac{(\nu(C_i))^{k_i}}{k_i!}\right).
  \end{multline}
  If $\nu(\d z)=\lb \d z$, $\pi_\nu$ is called the homogeneous Poisson
  point process with intensity parameter $\lb$ on $\R^d$.
\end{definition}
Actually, words for words, this definition does not need that $E$ is
an $\R^d$-like space. For the mathematical details to work, it is
sufficient to have $E$ a metric space with some weak topological
properties. It is a useful point of view since it is often interesting
to add some information to the location of users when these are
represented by the realization of Poisson point process. For instance,
one may want to add to the position of a customer, the fading and/or
the shadowing he is experiencing, his traffic rate, etc. In the
simplest case, all these characteristics should be independent from
one user to the other and identically distributed. We then say that
they are \textsl{marks} of the Poisson process. Under the above
mentioned hypothesis of independence and identity in distribution, the
process whose particles are couples $(x,\, v)$ is still a Poisson
process on the product space $\R^d\times \Vspace$ where
$\Vspace$ is the space in which the marks ``live''. The
intensity measure of this process is the product of $\lambda \d x$
times the probability distribution of the marks, denoted by $d\Vdist(v)$.  Many quantities we
are longing to compute are expressible as a sum over the points of a
realization of a deterministic function:
\begin{equation*}
  F=\sum_{x\in \eta}f(x,m)
\end{equation*}
where $\eta$ is a realization of $\pi_\mu$. The calculations of the
different moments of such a functional turn to be known and resort to
the Bell polynomials. The complete Bell polynomials $B_n(a_1,...,a_n)$
are defined as follows:
\[\exp\left\{\sum_{n=1}^{\infty}\frac{a_n}{n!}\theta^n\right\}=\sum_{n=1}^{\infty}\frac{B_n(a_1,a_2,...,a_n)}{n!}\theta^n\]
for all $a_1,...,a_n$ and $\theta$ such that all above terms are
correctly defined.  The first four Bell complete polynomials are given
as:
\begin{eqnarray*}
  B_1(a_1)                        &=& a_1\\
  B_2(a_1,a_2)                    &=& a_1^2+a_2\\
  B_3(a_1,a_2,a_3)                    &=& a_1^3+3a_1a_2+a_3\\
  B_4(a_1,a_2,a_3,a_4)                    &=& a_1^4+4a_1^2a_2+4a_1a_3+3a_2^2+a_4\\
\end{eqnarray*}

\begin{theorem}[Generalized Campbell's formula]\label{theorem:moments}
  Let $\nu_\lambda(\d x,\, \d v)=\lambda \d x \otimes \d\Vdist(v)$ and $n$
  be an integer and assume that $f\in L^p(E,\nu_\lambda)$ for $p\ge
  n$.  The moments of $F$ are given by:
  \begin{equation*}
    \esp{}{F^n}=
    B_n\left(\int_Ef(z)\d \nu_\lambda(z),\,\cdots,\,\int_Ef^n(z)\d \nu_\lambda(z)\right)
  \end{equation*}
  where $z=(x,\, v)$.
\end{theorem}
Poisson point processes enjoy a lot of useful properties for thinning,
superposition and displacement which roughly say that whatever one of
these transformations we apply to a Poisson point process, the
resulting process is still a Poisson point process with a tractable
intensity measure (see
\cite{StochasticGeometryandWirelessNetworksVolumeI} for complete
references). In the forthcoming computations, we need a more recently
established property.
From \cite{Coutin:eu,DFMV:2012}, we have the following theorem.
\begin{theorem}
  \label{thm_myglauber:2}
  Let $\nu_\lambda(\d x,\, \d v)=\lambda \d x \otimes \d\Vdist(v)$ and
  $\pi_{\nu_\lambda}$ be a Poisson process of intensity measure $\nu_\lambda$.
  For some function $f$ sufficiently integrable, let
  $F(\eta)=\sum_{x\in \eta} f(x,\, v)$ and
  \begin{equation*}
    \tilde{F}(\eta)=\dfrac{F(\eta)-\int f(x,\, v)\nu_\lambda(\d x,\d v)}{\left({\int f(x,\, v)^2\nu_\lambda(\d x,\d v)}\right)^{1/2}}\cdotp
  \end{equation*}
  For any $p\ge 1$, let
  \begin{multline*}
    m(p,\,\lambda)=(\int_C f^2(x,\, v)\nu_\lambda(d x,\d v))^{-p/2}\\
    \times (\int_C |f(x,\, v)|^p\nu_\lambda(\d x,\d v).
  \end{multline*}
  Let $\mu$ be the standard Gaussian measure on $\R$ and $\mu_3$, the measure
  given by
  \begin{equation*}
    d\mu_3(x)=(1+\frac{1}{6}m(3,\, \lambda) H_3(x))\d\mu(x),
  \end{equation*}
  where $H_3(x)=8x^3-12x$ is the third Hermite polynomial.  Then,
  \begin{equation*}
    \sup_{\|\psi\|_{\mathcal C^3_b}\le 1}\left|\esp{}{\psi(\tilde{F})}-\int_\R
      \psi\d\mu_3\right|\le E_\lambda
  \end{equation*}
  where
  \begin{equation*}
    E_\lambda=\left(\frac{m(3,\, 1)^2}{6}+\frac{m(4,\, 1)}{9}\sqrt{\frac2\pi}\right)\lambda^{-1}.
  \end{equation*}
\end{theorem}
  This means that for $E_\lambda$ small, the distribution of
  $\tilde{F}$ is in some sense very close to $\mu_3$.

\section{System model}\label{ONOFF section Model}


For the application we have in mind, i.e. deployment of a cellular
network in low populated region, one can consider an isolated cell,
neglecting the interference from adjacent base stations. To simplify
the computations, we consider that the region covered by the base
station, hereafter called the cell and denoted by $C$, is circular of
radius $R$, centered at the base station $o$. The forthcoming analysis
can be extended to any bounded domain of coverage but the integrals
would have to be numerically evaluated. The terminals are identified
to a cloud of points, which we denote by $\eta$, whose elements are
the positions of each terminal in a domain larger than $C$.

The power consumed by the battery of the base station $o$ can be
divided into two parts:
\begin{itemize}
\item The power dedicated to transmit, receive, decode and encode the
  signal of any active user. The cumulative power over the whole
  configuration is then the sum over all terminals of the energy
  consumed for each one.
\item The power dedicated to broadcast messages. In order to guarantee
  that all active users receive these messages, the power must be such
  that the farthest user in the cell is within the reception range (if
  the system performs power control) or all the cell is within
  reception range (if the system does not perform power
  control). Thus, the power is a function of the maximum distance
  between the base station and the terminals or it is constant if
  power control is not performed.
\end{itemize}

For a very simple propagation model (without fading and shadowing),
the Shannon's formula states that for a receiver located at $x$, the
transmission rate is given by
\begin{equation*}
  W\log_2(1+P_el(x)),
\end{equation*}
where $W$ is the bandwidth, $P_e$ is the transmitted power and $l(x)$
is the path-loss function. This implies that in order to guarantee a
minimum rate at position $x$, $P_e$ must be proportional to $1/l(x)$.
Usual choices of path-loss functions are of the form
$l(x)=\norm{x}^{-\gamma}$ (singular path-loss model) or
$l(x)=(r_0\vee\norm{x})^{-\gamma}$ or
$l(x)=(1+\norm{x}^{-\gamma})^{-1}$. The forthcoming analysis does not
depend on a particular choice, so we keep it generic.  It follows that
for a user configuration $\eta$, the total consumed power, in presence
of power control, is given by:
\begin{multline*}
  \Pt(\eta) = \beta_A\sum_{x\in\eta\cap C, x \
    \text{active}}l^{-1}(x)+\beta_B\max_{x\in \eta\cap C} l^{-1}(x)
  \\:=\Pa(\eta) + \Pb(\eta),
\end{multline*}
where $\beta_A$ and $\beta_B$ are multiplicative factors defined
below. The subscript A stands for "additive" and B stands for
"broadcast". The term $\sum_{x\in \eta} l^{-1}(x)$ means that we add
over all points $x$ of the configuration $\eta$, the value of
$l^{-1}(x)$. Now, if the terminals are moving, we denote by $\eta_t$
the configuration at time $t$ which represents the locations of all
terminals at this instant. Since the energy is the integral of the
power over time, the total consumed energy between time $0$ and time
$T$ is given by
\begin{multline*}
  \Et := \Et(\eta,T) = \int_0^T \Pa(\eta_s)\d s + \int_0^T
  \Pb(\eta_s)\d s \\
  =\Ea(\eta,T)+\Eb(\eta,T).
\end{multline*}
We should also add a constant part for the energy associated to
operate the network but it doesn't alter the statistical aspects we
aim to analyze.

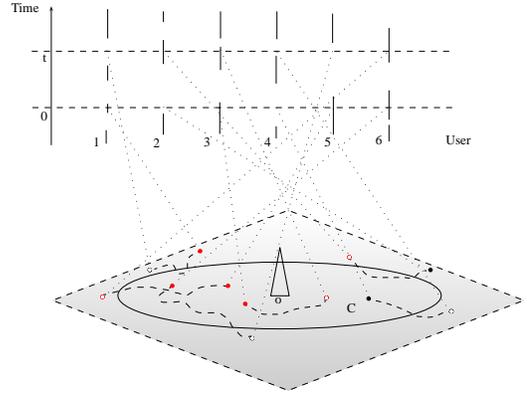
\begin{figure}
  \begin{center}
    \scalebox{0.5} 
    {
      \begin{pspicture}(0,-5.164152)(13.627476,5.205)
        \definecolor{color1882f}{rgb}{0.8,0.8,0.8}
        \definecolor{color1639}{rgb}{0.8,0.0,0.0}
        \psdiamond[linewidth=0.024,linestyle=dashed,dash=0.16cm
        0.16cm,dimen=outer,fillstyle=gradient,gradlines=2000,gradbegin=white,gradend=color1882f,gradmidpoint=1.0](7.3490624,-2.755)(6.3,2.41)
        \psellipse[linewidth=0.024,dimen=outer](7.1190624,-2.625)(4.31,0.9)
        \pstriangle[linewidth=0.024,dimen=outer](7.1290627,-2.645)(0.52,1.36)
        \pscustom[linewidth=0.024,linestyle=dashed,dash=0.16cm 0.16cm]
        { \newpath \moveto(4.2490625,-2.365) \lineto(4.2090626,-2.415)
          \curveto(4.1890626,-2.44)(4.1290627,-2.485)(4.0890627,-2.505)
          \curveto(4.0490627,-2.525)(3.9840624,-2.565)(3.9590626,-2.585)
          \curveto(3.9340625,-2.605)(3.9090624,-2.65)(3.9090624,-2.675)
          \curveto(3.9090624,-2.7)(3.9240625,-2.745)(3.9390626,-2.765)
          \curveto(3.9540625,-2.785)(3.9990625,-2.805)(4.0290623,-2.805)
          \curveto(4.0590625,-2.805)(4.1140623,-2.805)(4.1390624,-2.805)
          \curveto(4.1640625,-2.805)(4.2190623,-2.805)(4.2490625,-2.805)
          \curveto(4.2790623,-2.805)(4.3440623,-2.8)(4.3790627,-2.795)
          \curveto(4.4140625,-2.79)(4.4790626,-2.785)(4.5090623,-2.785)
          \curveto(4.5390625,-2.785)(4.5940623,-2.805)(4.6190624,-2.825)
          \curveto(4.6440625,-2.845)(4.6840625,-2.885)(4.6990623,-2.905)
          \curveto(4.7140627,-2.925)(4.7440624,-2.96)(4.7590623,-2.975)
          \curveto(4.7740626,-2.99)(4.8140626,-3.015)(4.8390627,-3.025)
          \curveto(4.8640623,-3.035)(4.9090624,-3.055)(4.9290624,-3.065)
          \curveto(4.9490623,-3.075)(4.9940624,-3.085)(5.0190625,-3.085)
          \curveto(5.0440626,-3.085)(5.1040626,-3.085)(5.1390624,-3.085)
          \curveto(5.1740627,-3.085)(5.2490625,-3.085)(5.2890625,-3.085)
          \curveto(5.3290625,-3.085)(5.3940625,-3.085)(5.4190626,-3.085)
          \curveto(5.4440627,-3.085)(5.4890623,-3.09)(5.5090623,-3.095)
          \curveto(5.5290623,-3.1)(5.5690627,-3.12)(5.5890627,-3.135)
          \curveto(5.6090627,-3.15)(5.6590624,-3.185)(5.6890626,-3.205)
          \curveto(5.7190623,-3.225)(5.7640624,-3.27)(5.7790623,-3.295)
          \curveto(5.7940626,-3.32)(5.8190627,-3.37)(5.8290625,-3.395)
          \curveto(5.8390627,-3.42)(5.8640623,-3.46)(5.8790627,-3.475)
          \curveto(5.8940625,-3.49)(5.9290624,-3.525)(5.9490623,-3.545)
          \curveto(5.9690623,-3.565)(6.0090623,-3.61)(6.0290623,-3.635)
          \curveto(6.0490627,-3.66)(6.0890627,-3.695)(6.1090627,-3.705)
          \curveto(6.1290627,-3.715)(6.1840625,-3.725)(6.2190623,-3.725)
          \curveto(6.2540627,-3.725)(6.3090625,-3.72)(6.3690624,-3.705)
        } \pscustom[linewidth=0.024,linestyle=dashed,dash=0.16cm
        0.16cm] { \newpath \moveto(5.0090623,-1.445)
          \lineto(4.9590626,-1.465)
          \curveto(4.9340625,-1.475)(4.8790627,-1.5)(4.8490624,-1.515)
          \curveto(4.8190627,-1.53)(4.7690625,-1.56)(4.7490625,-1.575)
          \curveto(4.7290626,-1.59)(4.6990623,-1.625)(4.6890626,-1.645)
          \curveto(4.6790624,-1.665)(4.6690626,-1.72)(4.6690626,-1.755)
          \curveto(4.6690626,-1.79)(4.6690626,-1.85)(4.6690626,-1.875)
          \curveto(4.6690626,-1.9)(4.6540623,-1.95)(4.6390624,-1.975)
          \curveto(4.6240625,-2.0)(4.5840626,-2.03)(4.5590625,-2.035)
          \curveto(4.5340624,-2.04)(4.4790626,-2.045)(4.4490623,-2.045)
          \curveto(4.4190626,-2.045)(4.3840623,-2.025)(4.3790627,-2.005)
          \curveto(4.3740625,-1.985)(4.3590627,-1.94)(4.3490624,-1.915)
          \curveto(4.3390627,-1.89)(4.3190627,-1.85)(4.3090625,-1.835)
          \curveto(4.2990627,-1.82)(4.2640624,-1.805)(4.2390623,-1.805)
          \curveto(4.2140627,-1.805)(4.1440625,-1.815)(4.0990624,-1.825)
          \curveto(4.0540624,-1.835)(3.9690626,-1.86)(3.9290626,-1.875)
          \curveto(3.8890624,-1.89)(3.8340626,-1.91)(3.7890625,-1.925)
        } \pscustom[linewidth=0.024,linestyle=dashed,dash=0.16cm
        0.16cm] { \newpath \moveto(9.469063,-2.725)
          \lineto(9.679063,-2.765)
          \curveto(9.784062,-2.785)(9.959063,-2.83)(10.029062,-2.855)
          \curveto(10.099063,-2.88)(10.224063,-2.925)(10.279062,-2.945)
          \curveto(10.334063,-2.965)(10.419063,-3.0)(10.449062,-3.015)
          \curveto(10.479062,-3.03)(10.5390625,-3.06)(10.569062,-3.075)
          \curveto(10.599063,-3.09)(10.659062,-3.125)(10.689062,-3.145)
          \curveto(10.719063,-3.165)(10.784062,-3.185)(10.819062,-3.185)
          \curveto(10.854062,-3.185)(10.9140625,-3.19)(10.939062,-3.195)
          \curveto(10.964063,-3.2)(11.024062,-3.205)(11.059063,-3.205)
          \curveto(11.094063,-3.205)(11.154062,-3.2)(11.179063,-3.195)
          \curveto(11.204062,-3.19)(11.249063,-3.18)(11.269062,-3.175)
          \curveto(11.2890625,-3.17)(11.334063,-3.15)(11.359062,-3.135)
          \curveto(11.384063,-3.12)(11.439062,-3.095)(11.469063,-3.085)
          \curveto(11.499063,-3.075)(11.569062,-3.06)(11.689062,-3.045)
        } \pscustom[linewidth=0.024,linestyle=dashed,dash=0.16cm
        0.16cm] { \newpath \moveto(11.029062,-1.965)
          \lineto(10.979062,-1.965)
          \curveto(10.954062,-1.965)(10.894062,-1.985)(10.859062,-2.005)
          \curveto(10.824062,-2.025)(10.774062,-2.065)(10.759063,-2.085)
          \curveto(10.744062,-2.105)(10.714063,-2.135)(10.699062,-2.145)
          \curveto(10.684063,-2.155)(10.639063,-2.165)(10.609062,-2.165)
          \curveto(10.579062,-2.165)(10.524062,-2.16)(10.499063,-2.155)
          \curveto(10.474063,-2.15)(10.419063,-2.145)(10.389063,-2.145)
          \curveto(10.359062,-2.145)(10.304063,-2.145)(10.279062,-2.145)
          \curveto(10.254063,-2.145)(10.184063,-2.145)(10.139063,-2.145)
          \curveto(10.094063,-2.145)(10.019062,-2.145)(9.989062,-2.145)
          \curveto(9.959063,-2.145)(9.849063,-2.145)(9.769062,-2.145)
          \curveto(9.689062,-2.145)(9.564062,-2.13)(9.519062,-2.115)
          \curveto(9.474063,-2.1)(9.404062,-2.055)(9.379063,-2.025)
          \curveto(9.354062,-1.995)(9.294063,-1.94)(9.259063,-1.915)
          \curveto(9.224063,-1.89)(9.179063,-1.845)(9.169063,-1.825)
          \curveto(9.159062,-1.805)(9.134063,-1.77)(9.119062,-1.755)
          \curveto(9.104062,-1.74)(9.054063,-1.705)(9.019062,-1.685)
          \curveto(8.984062,-1.665)(8.949062,-1.64)(8.949062,-1.625) }
        \pscustom[linewidth=0.024,linestyle=dashed,dash=0.16cm 0.16cm]
        { \newpath \moveto(6.1890626,-2.865) \lineto(6.2590623,-2.895)
          \curveto(6.2940626,-2.91)(6.3540626,-2.94)(6.3790627,-2.955)
          \curveto(6.4040623,-2.97)(6.4490623,-2.995)(6.4690623,-3.005)
          \curveto(6.4890623,-3.015)(6.5340624,-3.045)(6.5590625,-3.065)
          \curveto(6.5840626,-3.085)(6.6290627,-3.11)(6.6490626,-3.115)
          \curveto(6.6690626,-3.12)(6.7190623,-3.125)(6.7490625,-3.125)
          \curveto(6.7790623,-3.125)(6.8440623,-3.125)(6.8790627,-3.125)
          \curveto(6.9140625,-3.125)(7.0090623,-3.125)(7.0690627,-3.125)
          \curveto(7.1290627,-3.125)(7.2140627,-3.11)(7.2390623,-3.095)
          \curveto(7.2640624,-3.08)(7.3290625,-3.055)(7.3690624,-3.045)
          \curveto(7.4090624,-3.035)(7.4840627,-3.01)(7.5190625,-2.995)
          \curveto(7.5540624,-2.98)(7.7290626,-2.95)(7.8690624,-2.935)
          \curveto(8.009063,-2.92)(8.194062,-2.895)(8.239062,-2.885)
          \curveto(8.284062,-2.875)(8.329062,-2.84)(8.329062,-2.815)
          \curveto(8.329062,-2.79)(8.329062,-2.755)(8.329062,-2.725) }
        \pscustom[linewidth=0.024,linestyle=dashed,dash=0.16cm 0.16cm]
        { \newpath \moveto(5.7090626,-2.385) \lineto(5.6690626,-2.375)
          \curveto(5.6490626,-2.37)(5.6040626,-2.36)(5.5790625,-2.355)
          \curveto(5.5540624,-2.35)(5.4990625,-2.35)(5.4690623,-2.355)
          \curveto(5.4390626,-2.36)(5.3840623,-2.365)(5.3590627,-2.365)
          \curveto(5.3340626,-2.365)(5.2840624,-2.37)(5.2590623,-2.375)
          \curveto(5.2340627,-2.38)(5.1840625,-2.385)(5.1590624,-2.385)
          \curveto(5.1340623,-2.385)(5.0740623,-2.395)(5.0390625,-2.405)
          \curveto(5.0040627,-2.415)(4.9490623,-2.435)(4.9290624,-2.445)
          \curveto(4.9090624,-2.455)(4.8690624,-2.48)(4.8490624,-2.495)
          \curveto(4.8290625,-2.51)(4.7840624,-2.54)(4.7590623,-2.555)
          \curveto(4.7340627,-2.57)(4.6890626,-2.6)(4.6690626,-2.615)
          \curveto(4.6490626,-2.63)(4.6040626,-2.635)(4.5790625,-2.625)
          \curveto(4.5540624,-2.615)(4.5090623,-2.595)(4.4890623,-2.585)
          \curveto(4.4690623,-2.575)(4.4140625,-2.56)(4.3790627,-2.555)
          \curveto(4.3440623,-2.55)(4.2890625,-2.535)(4.2690625,-2.525)
          \curveto(4.2490625,-2.515)(4.1940627,-2.495)(4.1590624,-2.485)
          \curveto(4.1240625,-2.475)(4.0290623,-2.46)(3.9690626,-2.455)
          \curveto(3.9090624,-2.45)(3.8290625,-2.435)(3.8090625,-2.425)
          \curveto(3.7890625,-2.415)(3.7340624,-2.405)(3.6990626,-2.405)
          \curveto(3.6640625,-2.405)(3.6040626,-2.405)(3.5790625,-2.405)
          \curveto(3.5540626,-2.405)(3.4990625,-2.405)(3.4690626,-2.405)
          \curveto(3.4390626,-2.405)(3.3790624,-2.405)(3.3490624,-2.405)
          \curveto(3.3190625,-2.405)(3.2340624,-2.415)(3.1790626,-2.425)
          \curveto(3.1240625,-2.435)(3.0490625,-2.455)(3.0290625,-2.465)
          \curveto(3.0090625,-2.475)(2.9540625,-2.49)(2.9190626,-2.495)
          \curveto(2.8840625,-2.5)(2.7940626,-2.52)(2.7390625,-2.535)
          \curveto(2.6840625,-2.55)(2.5890625,-2.575)(2.5490625,-2.585)
          \curveto(2.5090625,-2.595)(2.4590626,-2.605)(2.4290626,-2.605)
        } \psdots[dotsize=0.12,linecolor=red](6.2090626,-2.845)
        \psdots[dotsize=0.12](11.129063,-1.945)
        \psdots[dotsize=0.12](9.489062,-2.705)
        \psdots[dotsize=0.12,linecolor=red](5.7490625,-2.365)
        \psdots[dotsize=0.12,linecolor=red](5.0090623,-1.445)
        \psdots[dotsize=0.12,linecolor=red](4.2690625,-2.365)
        \psdots[dotsize=0.12,fillstyle=solid,dotstyle=o](11.689062,-3.045)
        \psdots[dotsize=0.12,linecolor=red,fillstyle=solid,dotstyle=o](8.969063,-1.605)
        \psdots[dotsize=0.12,fillstyle=solid,dotstyle=o](3.6690626,-1.945)
        \psdots[dotsize=0.12,linecolor=red,fillstyle=solid,dotstyle=o](2.4090624,-2.665)
        \psdots[dotsize=0.12,fillstyle=solid,dotstyle=o](6.3890624,-3.765)
        \psdots[dotsize=0.12,linecolor=color1639,fillstyle=solid,dotstyle=o](8.369062,-2.685)
        \usefont{T1}{ptm}{m}{n} \rput(9.03625,-2.955){C}
        \usefont{T1}{ptm}{m}{n} \rput(7.0867186,-2.755){o}
        \psline[linewidth=0.024cm,arrowsize=0.05291667cm
        2.0,arrowlength=1.4,arrowinset=0.4]{->}(1.0490625,1.375)(1.0490625,5.075)
        \psline[linewidth=0.024cm,linestyle=dashed,dash=0.16cm
        0.16cm](0.5490625,2.375)(11.709063,2.375)
        \psline[linewidth=0.024cm,linestyle=dashed,dash=0.16cm
        0.16cm](0.5290625,3.875)(11.629063,3.875)
        \usefont{T1}{ptm}{m}{n} \rput(0.3584375,5.025){Time}
        \usefont{T1}{ptm}{m}{n} \rput(0.86609375,2.145){0}
        \usefont{T1}{ptm}{m}{n} \rput(0.8776562,3.685){t}
        \psline[linewidth=0.024cm](2.5490625,4.995)(2.5490625,4.215)
        \psline[linewidth=0.024cm](2.5490625,3.495)(2.5490625,3.095)
        \psline[linewidth=0.024cm](2.5490625,2.475)(2.5490625,2.235)
        \psline[linewidth=0.024cm](4.0290623,4.175)(4.0290623,3.535)
        \psline[linewidth=0.024cm](4.0290623,2.215)(4.0290623,1.655)
        \psline[linewidth=0.024cm](4.0290623,4.655)(4.0290623,4.935)
        \psline[linewidth=0.024cm](5.5490627,3.995)(5.5490627,3.375)
        \psline[linewidth=0.024cm](5.5290623,2.515)(5.5290623,1.675)
        \psline[linewidth=0.024cm](5.5490627,4.215)(5.5490627,4.935)
        \psline[linewidth=0.024cm](7.0490627,4.935)(7.0490627,4.195)
        \psline[linewidth=0.024cm](7.0290623,3.755)(7.0290623,3.015)
        \psline[linewidth=0.024cm](7.0290623,1.875)(7.0290623,1.555)
        \psline[linewidth=0.024cm](8.529062,4.875)(8.529062,4.095)
        \psline[linewidth=0.024cm](8.549063,2.675)(8.549063,1.655)
        \psline[linewidth=0.024cm](10.029062,4.495)(10.029062,3.575)
        \psline[linewidth=0.024cm](10.029062,2.835)(10.029062,2.055)
        \psline[linewidth=0.024cm](10.029062,1.915)(10.029062,1.475)
        \psline[linewidth=0.024cm,linestyle=dotted,dotsep=0.16cm](5.5090623,2.375)(6.1890626,-2.765)
        \psline[linewidth=0.024cm,linestyle=dotted,dotsep=0.16cm](5.5490627,3.915)(8.349063,-2.665)
        \psline[linewidth=0.024cm,linestyle=dotted,dotsep=0.16cm](8.549063,2.355)(5.7490625,-2.305)
        \psline[linewidth=0.024cm,linestyle=dotted,dotsep=0.16cm](8.529062,3.875)(6.4290624,-3.705)
        \psline[linewidth=0.024cm,linestyle=dotted,dotsep=0.16cm](9.989062,2.375)(4.3090625,-2.305)
        \psline[linewidth=0.024cm,linestyle=dotted,dotsep=0.16cm](9.969063,3.835)(2.4490626,-2.645)
        \psline[linewidth=0.024cm,linestyle=dotted,dotsep=0.16cm](2.5490625,2.375)(4.9890623,-1.405)
        \psline[linewidth=0.024cm,linestyle=dotted,dotsep=0.16cm](2.5290625,3.895)(3.6690626,-1.925)
        \psline[linewidth=0.024cm,linestyle=dotted,dotsep=0.16cm](4.0290623,2.395)(10.949062,-1.945)
        \psline[linewidth=0.024cm,linestyle=dotted,dotsep=0.16cm](4.0290623,3.875)(8.909062,-1.545)
        \psline[linewidth=0.024cm,linestyle=dotted,dotsep=0.16cm](7.0290623,2.375)(9.509063,-2.665)
        \psline[linewidth=0.024cm,linestyle=dotted,dotsep=0.16cm](7.0490627,3.855)(11.629063,-3.025)
        \usefont{T1}{ptm}{m}{n} \rput(11.869219,1.505){User}
        \usefont{T1}{ptm}{m}{n} \rput(2.2559376,1.465){1}
        \psline[linewidth=0.024cm](2.5090625,1.775)(2.5090625,1.415)
        \usefont{T1}{ptm}{m}{n} \rput(3.8476562,1.445){2}
        \usefont{T1}{ptm}{m}{n} \rput(5.1967187,1.465){3}
        \usefont{T1}{ptm}{m}{n} \rput(6.81,1.465){4}
        \usefont{T1}{ptm}{m}{n} \rput(8.398594,1.465){5}
        \usefont{T1}{ptm}{m}{n} \rput(9.764375,1.505){6}
      \end{pspicture}
    }
  \end{center}
  \caption{Illustration of the model, each user is associated with a
    ON-OFF process and a mobility process.}\label{fig: ON OFF PPP}
\end{figure}

For years, models for the locations of users in cellular networks were
left aside considering a sort of diffuse ether from which a density of
calls per unit of surface and unit of time would emerge. After
\cite{StochasticGeometryandWirelessNetworksVolumeI,StochasticGeometryandWirelessNetworksVolumeII},
we know how to represent users locations by a Poisson point
process. Note that according to the Mecke formula
\eqref{eq_myglauber:9}, the earlier fluid model can be viewed as a
space average of this refined description. As is, one cannot expect to
compute variances and higher order statistics from this model. We
hereby consider that terminals are initially located according to a
Poisson point process in the plane, of intensity $\lambda$: for two
disjoint bounded subsets of the plane, the random variables counting
the number of users in each subset are independent and Poisson
distributed with parameter $\lambda$ times the surface of the subset.
We enrich the Poisson point process description by adding traffic and
mobility characteristics. The traffic of the user initially located at
$x$, is an ON/OFF process, denoted by $A_x$, independent of the
position of the user. We assume, as usual, that all the traffic
processes of all users are independent and identically
distributed. Moreover, at the beginning of the time observation
window, they have all reached they stationary state (supposed to
exist). We denote by $\pion$ the probability for a given traffic
process to be in its ON phase at any given time. One simple example of
such a process is the exponential ON/OFF model, in which exponentially
distributed ON periods alternate with exponentially distributed OFF
periods. If we denote by $\muon$ and $\muoff$ the parameters of the
exponential distributions, then $\pion=\muoff/(\muon+\muoff)$. The
choice of a traffic model boils down to choose a probability measure
on the space ${\mathfrak T}$ of piece-wise, two valued, functions. We
denote by ${\mathcal T}$ this probability measure, hence ${\mathcal
  T}(\d a)$ is the probability to have a traffic process close to the
process $a$ and
\begin{equation*}
  \int H(a)\, \T(\d a)=\esp{}{H(A)},
\end{equation*}
means that we compute the mean value of a function $H$ with respect to
all possible values of the generic traffic process $A$.  We also
envision the impact of mobility on energy consumption. We just assume
that users move independently and are statistically indistinguishable:
If $M_x$ denote the movement of user initially located at $x$, so that
its position at time $t$ is $x+M_x(t)$, then we assume that the
collection of processes $(M_x,\, x\in \eta_0)$ are independent and
identically distributed. Besides the motionless situation where
$M_x(t)=o$ for any $t$ and any $x$, the simplest model is that
constant speed movement: $M_x(t)=v_xt$ where the vectors $(v_x,\, x\in
\eta_0)$ are independent and identically distributed over
$\R^2$. Choosing a mobility model boils down to determine a
probability measure $\M$ on the space $\mathfrak C$ of continuous
functions on $\R^2$, starting at $o$. Putting the pieces together
means that we consider a Poisson process of the product space
$\R^2\times {\mathfrak T}\times {\mathfrak C}$ with intensity $\lambda
\d x \otimes {\mathcal T}(\d a)\otimes \M(\d m)$. In plain words, this
means that a user, say located at $x$, is equipped with a traffic
process $A_x$ and a mobility process $M_x$ such that all these
processes are independent and identically distributed among all users.

Moreover, the so-called Mecke formula stands that
\begin{multline}\label{eq_myglauber:9}
  \esp{}{\sum_{x\in \eta} \zeta(x,\, A_x,\, M_x)}\\
  =\iint \zeta(x,\, a,\, m) \lambda \d x \, {\mathcal T}(\d a)\, \M(\d
  m),
\end{multline}
for any bounded function $\zeta$. The configuration of users at
time~$t$ is
\begin{eqnarray*}
  \eta_t=\sum_{x\in \eta_0}\delta_{x+M_x(t)},
\end{eqnarray*}
while the configuration of active users is
\begin{eqnarray*}
  \sum_{x\in \eta_0}A_x(t)\, \delta_{x+M_x(t)}.
\end{eqnarray*}
In particular, the additive part of consumed energy can be rewritten
as:
\begin{equation}\label{eq_myglauber:10}
  \Ea(\eta,T) = \sum_{x\in \eta_0}\int_0^TA_x(t)l^{-1}(x+M_x(t))\1_{x+M_x(t)\in C}\d t
\end{equation}
and the broadcast part is:
\begin{equation}\label{eq_myglauber:11}
  \Eb(\eta,T) = \int_0^T\max_{x\in \eta_t\cap C}l^{-1}(x+M_x(t))\d t.
\end{equation}
\section{Motionless users}
\label{sec:motionless-users}
When users do not move, from \eqref{eq_myglauber:10}, we get 
\begin{equation*}
  \Ea(\eta,T) = \beta_A\sum_{x\in \eta_0}\1_{x\in C}\left(\int_0^TA_x(t)\d t\right) \ l^{-1}(x),
\end{equation*}
so that $\Ea(\eta,T)$ appears as a shot noise process.  In view of
Theorem \ref{theorem:moments}, we can compute easily the moments of
any order of the additive part.
\begin{theorem}\label{theorem: Moments of Ea no mobi On OFF}
  For motionless users, the moments of $\Ea(\eta,T)$ are given by:
  \begin{equation*}
    \esp{}{ \Ea(\eta,T)^n}=B_n(\alpha_1,\cdots,\alpha_n)
  \end{equation*}
  where
  \begin{equation*}
    \m{k}{A,T}=\esp{\T}{\left(\int_0^T A(s)\d s\right)^k} \text{ for } k\ge 2.
  \end{equation*} and
  \begin{equation*}
    \alpha_k=\lambda \beta_A^k\m{k}{A,T}\int_C |l^{-1}(x)|^k\d x.
  \end{equation*}
  In particular, for the singular path-loss model,
  \begin{equation}\label{equ: Esp of Ea ON OFF}
    \Esp{\Ea(\eta,\,T)} = \frac{2\beta_A}{\gamma+2}\,\rho\,
    R^\gamma\ T
  \end{equation}
  where $\Nm=\lambda \pi R^2$ is the mean number of terminals into the
  cell and $\rho=\Nm\, \pion$ is the mean number of active customers
  in the cell $C$ of radius $R$.
\end{theorem}
We now show how to determine $\beta_A$. If $P_e$ denotes the power
emitted by a mobile located at $x$ from the base station, since we do
not take into account interference, shadowing and fading, the received
power at the base station is
\begin{math}
  P_r=P_e\, K\, l(x)
\end{math}
where $K=(c/4\pi f d_{ref})^2d_{ref}^\gamma$, $f$ is the frequency of
the radio transmission, $c$ is the light celerity and $d_{ref}$ is the
so-called reference distance.  Since the base station can detect a
signal of power greater than $\pminMS$, this requires that $P_e$ is
greater than $\pminMS/K l^{-1}(x)$. One can consider that the same
considerations hold for the downlink channel so that $\beta_A=2
\pminMS/K$.  In practical situation, $\pminMS$ is of the order of
$10^{-9}$ mW and $f$ around $2$ GHz, hence $\beta_A$ varies between
$2.10^{-10}$ for $\gamma=5$ to $2.10^{-8}$ for $\gamma=3$.

\begin{figure}[!ht]
  \includegraphics*[width=\columnwidth]{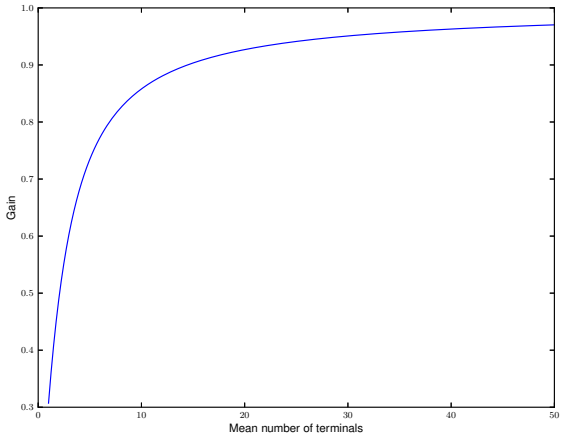}
  \caption{The energy gain with power control.}
  \label{fig_myglauber:control}
\end{figure}

On the other hand, without power control, let $P'_r$ the power
sufficient to ensure a reception at any point of the cell. If we
denote by $\pminBS$ the minimum power for the beacon to be detected by
a mobile at distance $R$ from the base station, we should have
\begin{equation*}
  \frac{K P'_r}{R^\gamma}\ge \pminBS.
\end{equation*}
This amounts to say that we can take $\beta_B=\pminBS /K$. Usually,
$\pminBS$ is around $10^{-8}$ mW. In the usual frequency bands, we
obtain $\beta_B$ varying from $10^{-9}$ for $\gamma=5$ to $10^{-7}$
for $\gamma=3$.  If there is no power control, the energy consumed for
the beacon is thus equal to
\begin{equation}
  \label{eq_myglauber:3}
  \Eb^0(\eta, \, T)=  \beta_B R^\gamma T.
\end{equation}
If power control is used, the power should be adjusted for the
farthest terminal to be able to receive the beacon signal:
\begin{equation*}
  \frac{\pminBS}{K}\max_{x\in \eta\cap C} l^{-1}(x)=\beta_B
  \max_{x\in \eta\cap C} l^{-1}(x).
\end{equation*}
Hence, in absence of movement,
\begin{equation*}
  \Eb(\eta,T) =\beta_B  T\ \max_{x\in \eta\cap C}l^{-1}(x).
\end{equation*}
Since all the usual path-loss functions depend only on the distance
between $x$ and $o$, let $L$ be defined as $L(\|x\|)=l(x)$. From the
remark that
\begin{equation*}
  (  \max_{x\in \eta\cap C} \|x\|\le u)=(\eta(B(o,R)\backslash B(o,\, u))=0),
\end{equation*}
it follows from \eqref{eq_myglauber:4} that the random variable
$\delta=\max_{x\in \eta\cap C} \|x\|$ has probability density
function:
\begin{equation*}
  f_\delta(u)=2 \lambda \pi e^{-\lambda \pi R^2}\, u\, e^{\lambda \pi u^2}.
\end{equation*}
The next result follows.
\begin{theorem}
  \label{thm_myglauber:1}
  For motionless users, with power control, the consumed energy to
  maintain the beacon signal, denoted by $\Eb^p$ has moments given by:
  \begin{multline}
    \label{eq_myglauber:5}
    \esp{}{(\Eb^p(\eta,T))^k}=\left(\beta_B T\right)^k \ \int_0^R
    (L^{-1}(u))^k \, f_\delta(u)\d u,\\
    \text{ for any } k\ge 1.
  \end{multline}
  For the singular path-loss function, we obtain
  \begin{equation*}
    \esp{}{\Eb(\eta,T)}=e^{-\Nm}\Nm^{-\gamma/2}\int_0^{\Nm}v^{\gamma/2}e^{v}\d
    v\ \Eb^0(\eta, \, T).
  \end{equation*}
\end{theorem}
Thus the gain of power control does depend only on the mean number of
terminals whatever the radius, be it a few meters or some kilometers.
Figure \ref{fig_myglauber:control} shows that, as expected, the gain
is higher for lower load.

In view of \eqref{equ: Esp of Ea ON OFF} and \eqref{eq_myglauber:3},
we have
\begin{equation}\label{eq_myglauber:2}
  \kappa:=\frac{  \esp{}{\Ea(\eta,\, T)}}{\Eb^0(\eta,\,
    T)}=\frac{2}{\gamma+2}\, \frac{\rho
    \beta_A}{\beta_B}=\frac{2}{\gamma+2} \frac{\rho\pminMS}{\pminBS}\cdotp
\end{equation}
Since $\rho$, the mean number of active users is of the order of $10$
to $50$, in view of the values of $\beta_A$ and $\beta_B$,
\eqref{eq_myglauber:2} says that $\kappa$ depends essentially of the
ratio between $\pminMS$ and $\pminBS$. For the values we considered,
this means that $\Ea$ and $\Eb$ are of the same order of
magnitude. 

\section{Impact of mobility}\label{section: Onoffmobility}
For the sake of simplicity for the proofs, we now assume that $\Eb$ is
constant, equal to $\Eb^0$ given by \eqref{eq_myglauber:3}. We now
evaluate the impact of mobility on energy consumption.

According to the displacement theorem for Poisson processes, we known
that for each time $t$, the point process $\eta^\M_t=(x+M_x(t),\, x\in
\eta_0)$ is a still a Poisson point process with intensity $\lambda \d
x$. It follows that the expectation of consumed energy does not depend
on the mobility model and is equal to the value for motionless users.

However, for higher order moments, the correlations between positions
at different instants are to be taken into account so that the
variance and other moments are different for truly mobile users.  Let
\begin{multline*}
  \digamma_n^M(f,T)=\\\int_{\R^2}\esp{\T,\,\M}{\puissance{\int_0^Tf(x+M(t))A(t)\1_{x+M(t)\in
        C}\d t}{n}}\d x.
\end{multline*}
The following theorem uses the same techniques of proof as before but
in a more involved fashion.
\begin{theorem}\label{theorem: OnoffMoReduce}
  The moments of $\Ea(\eta,T)$ with mobility are given by
  \begin{multline*}
    \esp{\T}{\Ea(\eta^\M,T)^n}=\\
    B_n(\lb\digamma_{1}^M(\beta_A\,
    l^{-1},T),\lb\digamma_{2}^M(\beta_A\,
    l^{-1},T),\cdots,\lb\digamma_{n}^M(\beta_A\, l^{-1},T)).
  \end{multline*}
  It follows that mobility reduces moments of $\Ea$; i.e
  \begin{equation}\label{eq_myglauber:1}
    \esp{\T,\, \M}{\Ea(\eta,T)^n}\le  \esp{\T,\, 0}{\Ea(\eta,T)^n}.
  \end{equation}
\end{theorem}
\begin{IEEEproof}
  The first part of the proof follows from
  Theorem~\ref{theorem:moments}.  Let $f(x)=\beta_A\,l(x)\,1_{x\in
    C}$. Since the Lebesgue measure is translation invariant,
  \begin{equation*}
    \int_{\R^2} f(x+y)\d x=\int_{\R^2} f(x)\d x.
  \end{equation*}
  If we combine that remark with H\"older inequality, we get that for
  $y_1,\cdots,y_n$ in $\R^2$,
  \begin{align}
    \label{eq:1}
    \int_{\R^2} \prod_{j=1}^n f(x+y_j)\d x & \le \prod_{j=1}^n (
    \int_{\R^2} f(x+y_j)^n\d x)^{1/n}\notag\\
    &=\int_{\R^2} f(x)^n\d x.
  \end{align}
  For the sake of presentation, we denote by $\d t$ the product
  measure $\otimes_{j=1}^n \d t_j$ According to \eqref{eq:1}, we get
  \begin{multline*}
    \digamma_n^M(f,T) \\
    \begin{aligned}
      & = \int_{\R^2}\int_{[0,\, T]^n}
      \esp{\T,\M}{\prod_{i=1}^n\left(f(x+M(t_i))A(t_i)\right)}\d
      t \d x\\
      & =\int_{\R^2}\int_{[0,\, T]^n} \esp{\M}{\prod_{i=1}^nf(x+M(t_i))}\esp{\T}{\prod_{i=1}^nA(t_i)}\d t\d x\\
      &= \int_{[0,\, T]^n}\esp{\T}{\prod_{i=1}^nA(t_i)} \int_{\R^2}\esp{\M}{\prod_{i=1}^nf(x+M(t_i))}\d x\d t \\
      &\leq \int_{[0,\, T]^n}\Esp{\prod_{i=1}^nA(t_i)}\int_{\R^d}f(x)^n\d x\d t \\
      &= \Esp{\puissance{\int_0^TA(t)\d t}{n}}\, \int_{\R^2}f(x)^n\d x\\
      &= \m{n}{A,T}\int_{\R^2}f(x)^n\d x.
    \end{aligned}
  \end{multline*}
  Since the coefficients of Bell polynomials are non-negative,
  \eqref{eq_myglauber:1} follows.
\end{IEEEproof}
We now study the limiting variance when the speed of particles is
large (for instance, terminals in a high speed train). We say that a
movement distribution $\M$ has property $\textbf{T}$ whenever
$\P(M(s)=M(t))=0$ for all $s\neq t$. We denote by $M^\epsilon$ the
accelerated version of $M$ (and $\M^\epsilon$ the corresponding
probability distribution on ${\mathfrak C}$):
$M^\epsilon(t)=M(t)/\epsilon$. The full proof of the next result is
given in \cite{Vu12Thesis}.
\begin{theorem}
  If $M$ has the property $\textbf{T}$ then in high mobility regime,
  the variance of $\Ea$ tends to $0$, i.e
  \begin{equation*}
    \Var{\Ea(\eta^{\M^\epsilon},T)}\rightarrow0 \ \text{as} \ \epsilon\rightarrow 0.
  \end{equation*}
\end{theorem}
The above results say that, when users move the total consumed energy
by a base station does not change \emph{in average}, and the moments
of the additive part are reduced. Moreover, when users move very fast,
the consumed energy during a time period is almost constant. We can
see this as a consequence of weak central limit theorem. When users
move faster, the configuration of users is more mixing during a same
period of time, thus converge faster to the mean.


\section{Applications}\label{sec: Application}

\subsection{Dimensioning optimal cell size}

Consider an operator aiming to design the optimal cell radius $R$ to
cover a region of total area $S\subset\R^2$. We assume that the cells
are circular. The average total cost of the network is assumed to be
the sum of the operating cost during the life time of the network (say
$T$) and the cost of facilities (base stations). We assume a fixed
deployment cost for any base station regardless of its transmission
range. The number of base stations is then roughly equal to $S \,
R^{-2}$ so the installation cost of base stations is ${c_1\,
  S}{R^{-2}}$ with $c_1>0$. The operating cost is assumed to be
proportional to the consumed energy.

We assume that $l$ is the singular path-loss function,
i.e. $l(x)=\|{x}\|^{-\gamma}$. From the previous results, the mean energy
consumed by the network during its operating time is:
\begin{equation*}
  \frac{S}{R^2}(1+\kappa)\beta_B R^\gamma T.
\end{equation*}
This is an increasing function of $R$, which means that small cell
systems will consume less energy than larger cell systems. The average
total cost for the network is then
\begin{equation}\label{eq_myglauber:6}
  \Cost (R)      = S(1+\kappa)\beta_B T
  R^{\gamma-2}+\frac{c_1\,
    S}{R^2}.
\end{equation}
Note that $\kappa$ implicitly depends on $R$, as the larger the cell,
the higher the mean number of active customers. Equation
\eqref{eq_myglauber:6} shows that there are two antagonist trends:
large radius cells minimize the deployment cost whereas they increase
the operating cost.

If we keep $\kappa$ constant, i.e. we may have larger cell providing
that the mean number of customers per unit of surface is decreasing in
such a way $\lambda R^2$ is constant; the optimization problem has a
solution obtained by differentiation:
\begin{equation*}
  R_{\text{opt}}=\left(\frac{2c_1}{(\gamma+2)(1+\kappa) \beta_B T}\right)^{1/\gamma}\cdotp
\end{equation*}
As expected, the optimal radius depends heavily on the value of
$\gamma$ which is linked to the density of obstacles in the path of
radio waves.
\subsection{Dimensioning cell battery}
\begin{figure}[!ht]
  \includegraphics*[width=\columnwidth]{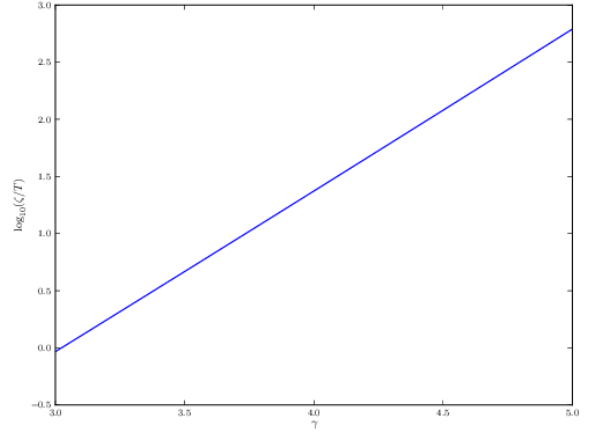}
  \caption{Variations of $\log_{10}(\zeta/T)$ with respect to
    $\gamma$.}
  \label{fig_myglauber:zeta}
\end{figure}

The proposed model can be used to dimension sites that do not have
access to power supply facilities. In this situation, operators have
to replace or reload base station's battery at each period $T$. We
want to determine the energy level $\alpha$ of battery so that the
probability of running out of energy before replacement (or reloading)
be smaller than some given threshold $\epsilon$. We use results
derived in the previous sections to find $\alpha$.  The problem is to
find $\alpha$ such that:
\begin{eqnarray*}
  \Proba{\Et(\eta^M,T)> \alpha} < \epsilon.
\end{eqnarray*}
In order to simplify the problem, we assume $\Eb$ to be constant equal
to
$K_B=\beta_B R^\gamma T$ and that the users are motionless. In view of
Theorem \ref{thm_myglauber:1}, it may be thought as a pessimistic
point of view without the economy due to power control. Hence,
$\Et(\eta^M,T)=\Ea(\eta^M,T)+K_B$.  One could resort to the
Bienaym\'e-Tcebycev inequality and use Theorem~\ref{theorem: Moments
  of Ea no mobi On OFF} to bound the variance of $\Et$ but this
approach is known to give imprecise results. Otherwise, one can use
the Gaussian approximation of Theorem \ref{thm_myglauber:2}. We can
apply this theorem to the function
$$f(x,a)=\beta_A\ l^{-1}(x)\int_0^Ta(t)\d t,$$
and hence,
$$\int_{C}\int_{\T}f(x,a)^k\nu_\lambda(\d x,\d
a)=\m{k}{A,T}\frac{(\beta_A R^{\gamma})^k}{\gamma k/2 +1}\, \Nm.$$ It
follows that for any $k\ge 3$,
\begin{equation*}
  m(k,\, \lambda)=\frac{\gamma+1}{\gamma
    k/2+1}\frac{\m{k}{A,T}}{\m{2}{A,T}^{k/2}}\, \Nm^{1-k/2}.
\end{equation*}
Since the process $A$ is supposed to be ergodic, it is well known that
$\m{k}{A,T}\sim (\pion\, T)^k$ as $T$ goes to infinity or at least if
$T$ is large compared to the cycle duration of $A$, i.e. the time
between two successive communications plus the length of a
communication. Since that is usually so, we get:
\begin{equation*}
  m(k,\, \lambda)=\frac{\gamma+1}{\gamma k/2+1}\,\Nm^{1-k/2}.
\end{equation*}
Note that $ m(k,\, \lambda)$ depends weakly on the geometric
properties of the domain which are summarized by $\gamma$ and on the
traffic pattern but mainly on the mean number of customers in the
cell.  The procedure is then the following: We first verify that
$E_\lambda$ is negligible compared to $\epsilon$. Then, we solve the
equation in $\alpha$,
\begin{equation*}
  \mu_3([\alpha,\, +\infty))= \epsilon
\end{equation*}
and take
\begin{align}
  \zeta &= \m{1}{A,T}\frac{\beta_A R^{\gamma}}{\gamma /2 +1}\,
  \Nm+\alpha \sqrt{\m{2}{A,T}}\frac{\beta_A R^{\gamma}}{\sqrt{\gamma
      +1}}\,\sqrt{\Nm}\notag\\
  &\sim\left(\frac{1}{\gamma /2 +1}+ \frac{\alpha}{\sqrt{\gamma
        +1}\sqrt{\Nm}}\right)\, \beta_A\rho R^\gamma T.
  \label{eq_myglauber:8}
\end{align}
With this procedure, we get a simple way to determine the threshold
$\zeta$ which guarantees that the real value of $
\Proba{\Et(\eta^M,T)> \alpha}$ is smaller than $\epsilon$. In
Figure~\ref{fig_myglauber:zeta}, we analyze the variations of $\zeta$
with respect to $\gamma$. Once again, we see that the provisioning of
resources, i.e. energy for the time being, is exponentially dependent
of $\gamma$, the path-loss exponent.

\section{Conclusion}
\label{sec:conclusion}

We have shown how to model energy consumption in a cellular network
taking into account both communication and signaling traffic. The
closed form formulas we obtained may be used for several purposes
mainly in order to dimension cells or batteries under energy
constraints. We pointed out the great importance of the domain
geometry which is summarized by the path-loss parameter $\gamma$.

\IEEEtriggeratref{3}


%
\end{document}

%% file: macro.tex
\def\argmax{\operatornamewithlimits{arg\,max}}
\def\argmin{\operatornamewithlimits{arg\,min}}
\newcommand{\dtc}{\ensuremath{\operatorname{d_{tc}}}}

\newtheorem{conjecture}{Conjecture}
\newtheorem{theorem}{Theorem}
\newtheorem{lemma}[theorem]{Lemma}
\newtheorem{remark}{Remark}
\newtheorem{claim}{Claim}
\newtheorem{observations}{Observation}
\def\CLf{C\`{a}dl\`{a}g }
\def\M{M}
\def\CauSw{Cauchy$-$Schwarz }
\def\PM{\P_M}
\def\car{{\mathbf 1}}
\def\N{{\mathbf N}}
\def\C{{\mathbf C}}
\def\R{{\mathbf R}}
\def\E{{\mathbf E}}
\def\S{{\mathbf S}}
\def\P{{\mathbf P}}
\def\Nm{{\mathfrak n}}
\def\Vspace{{\mathfrak V}}
\def\Vdist{{\mathcal V}}
\def\d{d}
\def\Esp#1{{\mathbf E}\left[#1\right]}
\def\Espp#1#2{{\mathbf E}_{#1}\left[#2\right]}
\def\Cov#1#2{{\mathbf C}\left[#1,#2\right]}
\def\Var#1{{\mathbf {V}}\left[#1\right]}
\def\mm#1#2{{\mathbf m}_{#1}\left[#2\right]}
\def\m#1#2{\mathbf{m}_{#1}\left(#2\right)}
\def\cm#1#2{{\mathbf c}_{#1}\left[#2\right]}
\def\pminBS{P_{\min}^{\text{b}}}
\def\pminMS{P_{\min}^{\text{r}}}
\newcommand\lip{{\text{Lip}}}
\newtheorem{definition}{Definition}

\newtheorem{proposition}[theorem]{Proposition}
\newtheorem{assumption}{Assumption}

\newtheorem{corollary}{Corollary}

\renewcommand{\d}{\operatorname{\, d}}
\def \bangle{ \atopwithdelims \langle \rangle}
\def\indicater#1{\mathbf{1}_{\{#1\}}}
\def\dd{\text{d}}
\def\Lb{\Lambda}
\def\x{x}
\def\W{W}
\def\1{1}
\def\lb{\lambda}
\def\lB{\overline{l}}
\def\B{\mathbf{B}}
\def\BBar{\overline{\mathbf{B}}}
\def\CBar{\overline{C}}
\def\indicator{T}
\def\E{\mathbf{E}}
\def\Ek{\mathbf{E}}
\def\dom{\text{dom}}
\def\var{\mathbf{var}}
\def\cov{\mathbf{cov}}
\def\Proba#1{\mathbf{P}\left(#1\right)}
\def\P{\mathbf{P}}
\def\R{\mathbb{R}}
\def\Pa{P_A}
\def\Cr{C(r)}
\def\CrBar{\overline{C}(r)}
\def\Pb{P_B}
\def\Pt{P}
\def\Pg{P_G}
\def\F{F}
\def\S{S_e}
\def\FB{\overline{F}}
\def\phiB{\overline{\phi}}
\def\psiB{\overline{\psi}}
\def\norm#1{\left\vert #1\right\vert}
\def\Norm#1{\left\Vert #1\right\Vert_\infty}
\def\puissance#1#2{\left( #1\right)^#2}
\newcommand\esp[2]{{\mathbf E}_{#1}\left[{#2}\right]}
\def\Cost{\text{Cost}}

\def\PPP{Poisson point process}
\def\Ea{J_A}
\def\EaM{J_{A}^{\M}}
\def\Eb{J_B}
\def\Eg{J_G}
\def\EgM{J_G^{\M}}
\def\EgMep{J_G^{\M/\epsilon}}
\def\EbM{J_{B}^{\M}}
\def\Et{J_T}
\def\EtM{J^{{\M}}}
\def\EtLT{J(N^{\Lb},T)}
\def\EaLT{J_A(N^{\Lb},T)}
\def\EbLT{J_B(N^{\Lb},T)}
\def\VNtLbI{\Vert  N^{\Lb}_t\Vert_{\infty}}
\def\VNtLbIM{\Vert  N^{\Lb}_{tM}\Vert_{\infty}}

\def\VNtoLbI{\Vert  N^{\Lb}_{t_1}\Vert_{\infty}}
\def\VNttLbI{\Vert  N^{\Lb}_{t_2}\Vert_{\infty}}
\def\VNtttoLbI{\Vert  N^{\Lb}_{t_2-t_1}\Vert_{\infty}}
\def\NtLb{N^{\Lb}_t}
\def\NtLbM{N^{\Lb,\M}_{t}}
\def\NtoLb{N^{\Lb}_{t_1}}
\def\NtoLbM{N^{\Lb,\M}_{t_1}}
\def\NttLb{N^{\Lb}_{t_2}}
\def\NttLbM{N^{\Lb,\M}_{t_2}}
\def\NtttoLb{N^{\Lb}_{t_2-t_1}}
\def\NLb{N^{\Lb}}
\def\NLbM{N^{\Lb,\M}}
\def\car{\mathbf{1}}
\def\om{\omega}
\def\Om{\Omega}
\def\NLbMobile{\underline{N}^{\Lambda}}
\def\NtLbMobile{\underline{N}^{\Lb}_t}
\def\Bc{\mathcal{B}}
\def\Fc{\mathcal{F}}
\def\Fbar{\overline{F}}
\def\Pc{\mathbf{P}}
\def\Lc{\mathcal{L}}

\def\Glauber{Generalized Glauber dynamic}
\def\GlauberLoss{Generalized Glauber dynamic with loss}
\def\GlauberMobility{Generalized Glauber dynamic with mobility}

\def\PoissonIPP{Poisson IPP model}
\def\PoissonONOFF{Poisson ON-OFF model}

\def\PoissonIPPMobility{Poisson IPP model with mobility}

\def\PoissonONOFFMobility{Poisson ON-OFF model with mobility}

\def\pion{\pi_{\text{ON}}}
\def\muon{\mu_{\text{ON}}}
\def\muoff{\mu_{\text{OFF}}}
\def\T{{\mathcal T}}
\def\M{{\mathcal M}}
\def\R{{\mathbb R}} 